\documentclass[12pt]{article}
\usepackage{amsmath,amsfonts,amsbsy,amssymb,amsthm,amstext,amscd}
\usepackage[psamsfonts]{eucal}
\usepackage{latexsym}
\usepackage{indentfirst,tabularx}

\def\centereps#1#2#3{\vskip#2\relax\centerline{\hbox to#1{\includegraphics{#3}\hfill}}}
\newcommand{\field}[1]{\mathbb{#1}}
\newcommand{\R}{\field{R}}
\newcommand{\h}{\field{H}}
\newcommand{\Z}{\field{Z}}

\newcommand{\abs}[1]{\lvert#1\rvert}
\newcommand{\norm}[1]{\lVert#1\rVert}
\theoremstyle{definition}

\newtheorem{thm}{Theorem}
\renewcommand{\u}{\textit}
\renewcommand{\-}{$\dfrac{\quad\enspace}{\quad}$}
\DeclareMathOperator{\cosec}{cosec}

\begin{document}
\begin{titlepage}
\setlength{\baselineskip}{16pt}
\begin{center}
\vskip 1in
{\large \textbf{AdS/CFT FOR NON-BOUNDARY MANIFOLDS}}
\vskip 1in
Brett McInnes
\break
Department of Mathematics\\
National University of Singapore\\
10, Kent Ridge Cresent, Singapore 119260\\
email : matmcinn@nus.edu.sg\\
\vskip 1in
ABSTRACT
\end{center}

In its Euclidean formulation, the AdS/CFT correspondence begins as a study of Yang-Mills conformal field theories on the sphere, $S^4$. It has been successfully extended, however, to $S^1\times S^3$ and to the torus $T^4$. It is natural to hope that it can be made to work for \u{any} manifold on which it is possible to define a stable Yang-Mills conformal field theory. We consider a possible classification of such manifolds, and show how to deal with the most obvious objection : the existence of manifolds which cannot be represented as boundaries. We confirm Witten's suggestion that this can be done with the help of a brane in the bulk.

\end{titlepage}
\setlength{\baselineskip}{20pt}

\renewcommand{\thesection}{\Roman{section}}
\section{\large INTRODUCTION}
In its original formulation, the AdS/CFT correspondence \cite{1} allows us to study Yang-Mills conformal field theories on $S^4$ in terms of an $S^5$ compactification of string theory on the hyperbolic space $H^5$. [We use the Euclidean approach throughout; $S^n$ is the $n$-sphere with the standard Riemmanian metric and conformal structure, and $H^n$ is the open ball $B^n$ endowed with the metric of constant sectional curvature equal to -1.] The transition from the gravitational theory in the ``bulk'' to a non-gravitational theory is effected by a geometric scheme in which $S^4$ appears as ``infinity'' for $H^5$. The obvious way \- but, we shall argue, \u{not} the only way \- to formulate this concretely is to regard $S^4$ as the \u{boundary} of the closed ball $\overline{B}^5$, after the manner of Penrose compactifications \cite{2} in general relativity.

It is generally agreed that AdS/CFT reflects some very deep property of these conformal field theories. If this is so, then surely the correspondence must work for manifolds other than $S^4$. This idea \cite{3} has led to some remarkable insights. For example, replacing $S^4$ by $S^1\times S^3$, we immediately note that there are at least \u{two} candidates for the bulk, $B^2\times S^3$ and $S^1\times B^4$. The conformal field theory partition function is then naturally defined by a sum of contributions from $B^2\times S^3$ and $S^1\times B^4$, which leads to a large $N$ phase transition related to black hole thermodynamics \cite{3}. Thus AdS/CFT can indeed be pushed beyond $S^4$. The obvious question now is : how far can it be pushed ? Are there manifolds for which AdS/CFT definitely does not work? If so, can one classify the manifolds for which it does? In short, what are the ``outer limits'' for AdS/CFT?

The ideal answer to the first of these questions would be : it works for every compact conformal manifold of dimension less than 11 on which a physically sensible CFT can be defined. Our objective in this work is not, of course, to ``prove'' such a grandiose assertion; our purpose, instead, is to formulate ``physically sensible'' in a precise way, and to answer the most obvious objection to this claim : how can it be true of a CFT defined on a compact manifold which is \u{not} the boundary of any manifold-with-boundary?

This question was already raised in \cite{3}, where it was suggested that an answer would involve introducing ``branes or stringy impurities of some kind'' into the bulk. We will argue here that \u{this is indeed precisely the correct answer}, and that AdS/CFT does have a chance of working even in this extreme case. Clearly, this will involve a more general formulation of the relationship between the bulk and ``infinity'' than is usually considered. We will see that recent important results on the geometry of ``infinity'' \cite{4}, \cite{5} can be interpreted physically as results on the nature of the matter content of the bulk.

We begin in section II by discussing a criterion, hinted at in \cite{3} and stated explicitly in \cite{6}, for a Yang-Mills CFT to be physically reasonable. This leads us, with the aid of the Kazdan-Warner classification \cite{7}, to a precise proposal as to the kinds of manifolds for which AdS/CFT should be expected to work. Among these are some manifolds which cannot be represented as boundaries, and so cannot be Penrose conformal infinity for any bulk. This leads us, in section III, to generalise the concept of conformal infinity in such a way that ``infinity'' is a hypersurface in a compact \u{manifold}, instead of a boundary of a manifold-with-boundary. Finally, in section IV, we use geometric techniques to prove that the bulk must, if ``infinity'' is not a boundary, contain some kind of ``brane or stringy impurity'', and to investigate the nature of these ``impurities''.

\section{\large THE STABILITY CONDITION FOR THE CFT}

It is pointed out in \cite{3} that the convergence of the path integral for the Yang-Mills CFT is non-trivial \- it depends on the geometry of the underlying manifold. Convergence is not a problem for $S^4$ with its standard conformal structure, but, as one moves away from this simplest case, one expects the good behaviour of the CFT to become increasingly questionable. We therefore need to know which properties of $S^4$ are essential and which are not. (Throughout this section, all manifolds have dimension $\geq 3$.)

As a conformal manifold (that is, a manifold on which one is given an equivalence class of conformally related metrics, as is appropriate for the study of conformal field theories), $S^4$ is distinguished by being \u{conformally flat}, and also by having a conformal structure represented by an Einstein metric. We wish to argue that neither of these properties is essential. Against conformal flatness we adduce the following evidence (apart from the fact that it is obviously an extremely severe restriction and would eliminate too many interesting manifolds). We shall see later that there is a compact manifold of the form $\R^4/\Gamma$, where $\Gamma$ is of course infinite but discrete, which is a boundary, but on which it is impossible to define a physically reasonable Yang-Mills CFT. \u{This is true for every conformal structure on} $\R^4/\Gamma$. And yet there is a conformally flat conformal structure on this manifold. Evidently the conformal flatness condition has little or no physical significance. Against the Einstein condition we have still stronger evidence, as follows. For any four-dimensional Einstein manifold $M$, the Euler characteristic is given by (\cite{8}. page 161)
\begin{equation}
\chi(M)=\dfrac{1}{8\pi^2}\int_M(\norm{U}^2+\norm{W}^2) dV,
\end{equation} 
where $U$ is the irreducible component of the curvature tensor determined by the scalar curvature, and $W$ is the Weyl tensor. Since $\chi(S^1\times S^3)=0$, an Einstein metric on $S^1\times S^3$ would necessarily be flat, an impossibility since $S^1\times S^3$ is not covered by $\R^4$. So there is no Einstein metric of any kind on $S^1\times S^3$, including non-product metrics. (This is just the simplest of several non-existence results of this kind; see \cite{9}.) Thus the Einstein condition would rule out $S^1\times S^3$ \- which, as we saw in the Introduction, is one of the most important examples of a manifold to which AdS/CFT \u{can} be extended.

Having abandoned constraints on the components of the curvature tensor determined by the Weyl and Ricci tensors, we turn naturally to its last remaining component, $U$. The proper criterion \cite{3},\cite{6} is as follows. Let $N^n$ be a compact manifold with a conformal structure $[g^N]$. The conformal Laplacian, obviously the physically relevant operator here, is \cite{10}
\begin{equation}
L_{g^N}=-\Delta_{g^N} +\dfrac{n-2}{4(n-1)}R(g^N),
\end{equation}
where $\Delta_{g^N}$ is the usual Laplacian, $n=\dim(N^n)$, and $R(g^N)$ is the scalar curvature. It is well known that $L_{g^N}$ is an elliptic operator with discrete real spectrum bounded from below. Let $\mu_1(g^N)$ be the first eigenvalue. Then the Yang-Mills CFT will be \u{stable} if $\mu_1(g^N)>0$, and sometimes when $\mu_1(g^N)=0$, but it is \u{unstable} if $\mu_1(g^N)<0$. This criterion can be stated more usefully as follows. By Schoen's solution \cite{11} of the Yamabe problem, $[g^N]$ contains a \u{Yamabe metric} $g^N_Y$ which, by definition, is such that $R(g^N_Y)$ is \u{constant} on $N^n$. This constant is given by
\begin{equation}
R(g^N_Y)=\frac{4(n-1)}{n-2}\mu_1(g^N_Y).
\end{equation}
The stability condition can therefore be expressed in terms of the sign of the scalar curvature. For example, $S^4$ with its usual metric has constant positive scalar curvature, so one can construct a stable CFT using \u{this} metric. However, $S^4$ also has \u{another} metric which \- surprisingly \- has constant \u{negative} scalar curvature. The CFT will of course be unstable if this metric is used. 

Now in fact Kazdan and Warner \cite{7} have given a classification of manifolds according to the behaviour of the scalar curvature. The following theorem is basic.
\begin{thm}[Kazdan-Warner]
Every compact connected manifold $M^n$, $n\geq 3$, falls into precisely one of the following  three classes.
\begin{enumerate}
\item $P$ : Every smooth function on $M^n$ is the scalar curvature of some metric on $M^n$.
\item $Z$ : A smooth function on $M^n$ is the scalar curvature of some metric if and only if it is either negative at some point or it is identically zero.
\item $N$ : A smooth function is the scalar curvature of some metric on $M^n$ if and only if it is negative at some point.
\end{enumerate}
\end{thm}
To see how to use this theorem, observe for example that since $S^4$ admits a metric of positive scalar curvature, it cannot be in $Z$ or $N$; hence it is in $P$; hence \u{every} smooth function on $S^4$ is the scalar curvature of some metric; hence indeed there is a metric on $S^4$ with constant negative scalar curvature, another with identically zero scalar curvature, and so on. Again, the torus $T^4$ accepts a flat metric, so it is not in $N$; since it is ``enlargeable'' (\cite{12}, page 306), it admits \u{no} metric of positive scalar curvature, so it cannot be in $P$; hence it is in $Z$. Finally, consider a compact 4-dimensional manifold which accepts a metric of constant negative sectional curvature. Such a manifold has the structure $\R^4/\Gamma$, for some discrete freely acting group $\Gamma$ with no subgroup of the form $\Z\oplus\Z\oplus\Z\oplus\Z$. ($\R^4/\Gamma$ is said to be homotopically atoroidal.) This manifold, too, is enlargeable, so it admits no metric of positive scalar curvature. Nor, however, does it admit a metric of zero scalar curvature, for such a metric on an enlargeable manifold must be flat, but $\R^4/\Gamma$ is not covered by $T^4$. Hence it is in $N$. This is the example mentioned earlier : its metric of constant negative sectional curvature is conformally flat (and so its Hirzebruch signature is zero; the signature being, in four dimensions, an isomorphism (\cite{12} page 92) from the oriented cobordism group to $\Z$, the manifold is a boundary) and yet \u{every} conformal class on $\R^4/\Gamma$ is represented by a Yamabe metric of negative scalar curvature, so conformal flatness certainly does not ensure satisfactory physical behaviour.

It is important to realise that the Kazdan-Warner classification is a classification of \u{manifolds} \- that is, the class to which a space belongs depends only on its topology and differentiable structure. (For this second point, note that $S^9$ with its usual differentiable structure belongs to $P$, but with a certain exotic differentiable structure \cite{13}, it belongs to $N$. The stability of these conformal field theories in nine (and ten) dimensions can therefore depend on the choice of differentiable structure.) Clearly it is not possible to define a stable CFT on a manifold in class $N$ \u{no matter which conformal structure we use}. The instability is \u{not} a geometric phenomenon for manifolds in class $N$; it is due to their differential topology. For manifolds in class $P$, by contrast, there is always a metric which makes the CFT stable, but there is also another metric which makes it unstable. (Notice that the Kazdan-Warner theorem implies that \u{every} compact manifold of dimension greater than two admits a metric of constant negative scalar curvature.) Hence, once it is known that a manifold is in $P$, the question of stability becomes a geometric question. (The reader should be aware that deciding the Kazdan-Warner class of a manifold can be non-trivial :  for example, simply connected six-dimensional Calabi-Yau manifolds are all in $P$, \u{not} $Z$.)

What of class $Z$? In all known examples, these manifolds behave like manifolds in class $P$. For example, $T^4$ with its flat metric admits a CFT which is perfectly well-behaved \cite{14}. The same appears to be true of manifolds and orbifolds of the form $T^4/\Delta$, where $\Delta$ is a finite group of isometries of $T^4$. (All such manifolds/orbifolds are in $Z$, since a metric of positive scalar curvature on $T^4/\Delta$ would pull back to a metric of positive scalar curvature on $T^4$.) Since $K3$ can be regarded as a resolution of a $T^4$ orbifold, we expect that the same is true in this case also. (That $K3$ is in $Z$ follows from the theorem of Lichnerowicz; see \cite{12}.) It is reasonable to conjecture that Kazdan-Warner class $Z$ is indeed like class $P$ : that is, for each manifold in $Z$, there is a conformal structure such that the CFT is stable, while there are of course other conformal structures such that the CFT is unstable.

Throughout this discussion, we have not assumed that ``infinity'', the compact manifold on which our CFT is defined, is \u{connected}. There is of course a well-defined stable CFT on $T^4+S^4$, the disjoint union of $T^4$ and $S^4$, though the two separate conformal field theories are very different and are decoupled.  Now there exists a non-compact five-dimensional spin manifold $M^5$ such that $T^4+S^4$ is the boundary of a connected manifold-with-boundary $\overline{M}^5$ having $M^5$ as interior. How can AdS/CFT work in this case? How can \u{one} interior be ``dual'' to \u{two} different, decoupled conformal field theories? This question was raised in \cite{4}. The simplest response to this paradox is just to declare that AdS/CFT should \u{not} be expected to work for disconnected boundaries (with an exception to be discussed below). In practical terms, this is of course a very minor limitation, since $T^4$ and $S^4$ are each boundaries, of $\overline{B}^2\times T^3$ and $\overline{B}^5$ respectively, so the separate conformal field theories can be studied by two applications of ordinary AdS/CFT. 

We can now state what we hope to be the full range of AdS/CFT. The claim is that AdS/CFT should work for \u{every compact connected infinity manifold} (or perhaps orbifold, etc) \u{of suitable dimension which does not belong to Kazdan-Warner} \u{class $N$}. By ``suitable dimension'' we mean simply that the bulk should be of dimension 10 or 11, or lower if there is a compactification.

So far, we have concentrated on the conditions to be satisfied by the differential topology and geometry of ``infinity'', without concerning ourselves with the details of the physical fields there. This is justifiable, in that the theory at ``infinity'' is avowedly non-gravitational. For precisely this reason, care should be exercised before imposing geometric conditions on the \u{bulk}. Often one assumes that the bulk is an Einstein manifold of Ricci curvature $-n$, but, while this is legitimate \cite{15} if ``infinity'' is $S^n$ and the conformal structure is not too far from the standard one, it should only be regarded as an approximation in other cases. In more general investigations \cite{16} the metric is only required to be asymptotically Einstein; no particular fall-off rate is assumed, and the metric is certainly \u{not} required to be complete \- indeed, in many applications it is definitely incomplete. Our attitude is that conditions on the metric in a \u{gravitational} theory should be dictated by the theory itself, not imposed externally. In short, we shall impose no requirements on the bulk metric.

As is observed already in \cite{3}, there is an obvious, strong objection to the idea that AdS/CFT works whenever the CFT is stable (that is, for connected compact manifolds of Kazdan-Warner classes $P$ and $Z$) : many manifolds are not boundaries. For example, no compact connected four-dimensional manifold is a boundary if its signature is not zero \cite{17}. We now deal with this objection.

\section{\large INFINITY IS JUST ANOTHER BRANE}

Let $\hat{M}^{n+1}$ be a compact $(n+1)$-dimensional manifold containing a smooth compact boundaryless hypersurface $N^n$. Fix a Riemannian metric $g^M$ on $M^{n+1}=\hat{M}^{n+1}-N^n$ and assume that there exists an ``infinity function'' f on $\hat{M}^{n+1}$. This is a smooth function which is positive on $M^{n+1}$ and vanishes to first order on $N^n$, such that $f^2g^M$ extends continuously to $N^n$. If such a function exists, then $N^n$ is ``infinitely far from'' points in $M^{n+1}$, and $g^M$ induces a conformal structure (not a Riemannian structure) on $N^n$. (Note that $M^{n+1}$ need not be complete.) In such a case, we shall say that $N^n$ is an \u{infinity hypersurface} for $\hat{M}^{n+1}$ with respect to $g^M$.

This definition is of course motivated by the formal definition \cite{15} of a Penrose conformal boundary, which is the more usual arena for AdS/CFT. Indeed, any compact manifold-with-boundary with the boundary ``at infinity'' can be re-interpreted in the above way : simply take two copies, and (adjusting the boundary orientation suitably) identify them along the boundary. The result will be a compact \u{manifold} with an infinity hypersurface at the former location of the boundary. One can also do this by beginning with distinct manifolds-with-boundary having diffeomorphic boundaries. For example, $\overline{B}^2\times S^3$ can be joined to $S^1\times \overline{B}^4$ along their common $S^1\times S^3$ boundary, and so the process of summing over distinct interiors \cite{3} can be implemented in a concrete way. Heuristically, there may well be advantages in dethroning ``infinity'' from its privileged position at the boundary, and thinking of it as ``just another brane'', one which happens to be infinitely far away; and certainly compact \u{manifolds} are preferable to manifolds-with-boundaries.

Clearly, AdS/CFT can be formulated in this language. Notice, however, that the infinity hypersurfaces obtained in this way have a special property : $N^n$ separates $M^{n+1}$ into disconnected pieces. By considering infinity hypersurfaces which do not have this effect, we obtain something new. The following family of examples is particularly enlightening.

Let $P^n$ be connected, compact, $n$-dimensional manifold with a Riemannian metric $g^P=g^P_{ij}dx^i\otimes dx^j$ and Ricci curvature $Ric(g^P)=R^P_{ij}dx^i\otimes dx^j$. Let $\hat{M}^{n+1}=S^1\times P^n$ with $S^1$ parametrised by $\theta$ running from $0$ to $2\pi$, and let $M^{n+1}$ be obtained from $\hat{M}^{n+1}$ by deleting all points with $\theta=0$. Define a metric $g^M$ on $M^{n+1}$ by
\begin{equation}
g^M=\cosec^2(\frac{\theta}{2})[\frac{1}{4}d\theta\otimes d\theta + g^P_{ij}dx^i\otimes dx^j].
\end{equation}
Here the function $f$ is $\sin(\dfrac{\theta}{2})$, which is positive in $(0,2\pi)$ and vanishes to first order at $\theta=0$, where there is a single copy of $P^n$. The infinity hypersurface does not separate $M^{n+1}$ into disconnected components. The Ricci tensor of this metric is, in an obvious notation,
\begin{align}
(R^M)^\theta_\theta&=-n,\\
(R^M)^i_j&=-n\delta^i_j+\sin^2(\frac{\theta}{2})[(R^P)^i_j+(n-1)\delta^i_j],
\end{align}
all other components being zero. We have expressed the Ricci tensor in $(1,1)$ form in order to be able to discuss \u{invariant} quantities, namely the \u{eigenvalue functions} of the Ricci curvature. (The $(0,2)$ components diverge near $\theta=0$, but this is merely a coordinate effect.) As $P^n$ is compact, the eigenvalue functions of $Ric(g^P)$ are bounded, and hence so are those of $Ric(g^M)$. For example, if $P^n$ is Ricci-flat, the eigenvalue functions of $Ric(g^M)$ are bounded above by $-1$ and below by their asymptotic value, $-n$.

\begin{figure}
\centereps{10cm}{5cm}{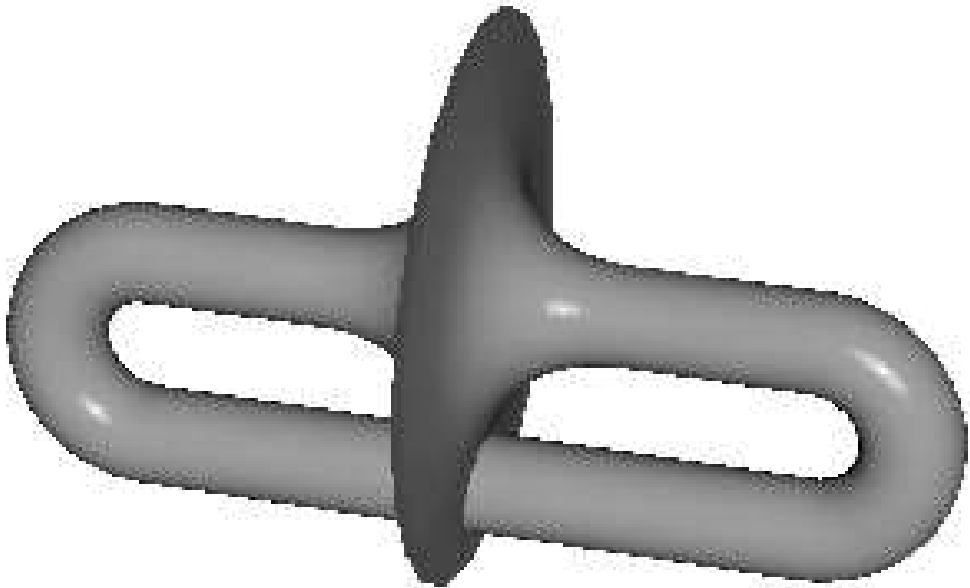}
\caption{Manifold with infinity hypersurface}
\end{figure}
The structure of this space (Figure 1) is clear : the infinity hypersurface is really one, connected copy of $P^n$, and the bulk $M^{n+1}$ is just an open submanifold. However, we can (``perversely'') re-interpret the structure of $M^{n+1}$as follows. Instead of the compact \u{manifold} $S^1\times P^n$, let us consider the compact manifold-with-boundary $[0,2\pi]\times P^n$ obtained by artificially distinguishing $\theta=2\pi$ from $\theta=0$. Now formula (4) still defines a Riemannian metric on the interior, $(0,2\pi)\times P^n$. If we insist on setting asunder what belongs together, we can regard $[0,2\pi]\times P^n$ as the Penrose compactification of $M^{n+1}$; the boundary now consists of \u{two} copies of $P^n$, one each at $\theta=0$ and $\theta=2\pi$. We can lend colour to this imposture by changing coordinates. Let $x$ be defined by
\begin{equation*}
\cosh(x)=\cosec(\frac{\theta}{2})
\end{equation*}
with $x\leq0$ for $\theta\leq\pi$, and $x\geq0$ for $\theta\geq\pi$. A short calculation reveals
\begin{equation}
g^M=dx\otimes dx + \cosh^2(x)g^P_{ij}dx^i\otimes dx^j,
\end{equation}
again suggesting two distinct ``infinities'', one at ``$x=-\infty$'', the other at ``$x=+\infty$''. (See \cite{8} page 268 and note that $g^M$ is Einstein if $Ric(g^P)=-(n-1)g^P$.)

In short, $M^{n+1}$ does not ``know'' whether it is an open submanifold of a compact \u{manifold}, or the interior of a compact manifold-with-boundary. However, the difference, from a physical point of view, is substantial. For the manifold-with-boundary interpretation leads us to a disconnected boundary, and so to the Witten-Yau paradox discussed previously. The ``infinity hypersurface'' interpretation is both more natural and more physically acceptable.

The metric (4) has no particular physical significance; it was chosen to make the above point. In general, we can take a compact manifold $\hat{M}^{n+1}$ with an infinity hypersurface $N^n$, and ``split'' it along $N^n$ to obtain a compact manifold-with-boundary having boundary components $N^n_1$, $N^n_2$, and so on, with each component diffeomorphic to $N^n$. We can then deform a conformal structure on $N^n_2$ through physically acceptable conformal structures to obtain a boundary component which is of course still diffeomorphic, but perhaps no longer isometric, to $N^n$. It would clearly be absurd to disallow the original manifold on the grounds that it can so artificially be brought into conflict with the Witten-Yau paradox. On the contrary, when we are presented with a metric like (7), representing the interior of a manifold-with-boundary with essentially identical boundary components, our first move should be to make the appropriate identifications and return to the more natural infinity hypersurface interpretation (corresponding to (4)). Of course, if the boundary components are not mutually diffeomorphic (or if they are, but their conformal structures can only be deformed to each other by passing through ``unstable'' conformal structures) then the paradox will lead to genuine difficulties.

These remarks bring us to our main application.

\section{\large MANIFOLDS WHICH ARE NOT BOUNDARIES}

Suppose that we wish to study a Yang-Mills CFT  on some compact manifold $N^n$ (such as a four-manifold with non-zero signature) which simply cannot be expressed as the boundary of some manifold-with-boundary. We now have a strategy : represent two copies of $N^n$ as a disconnected conformal boundary, and identify them to realise it as an infinity hypersurface in some compact manifold. The CFT should then be dual to some string theory in this ``bulk''. As we shall see, this can always be done; the only point at issue is whether the bulk admits a physically reasonable geometry.

When we say that a certain compact manifold $N^n$ is not a boundary, we mean ``not a boundary by itself''. It is always possible to find a compact manifold $Q^n$ such that $N^n+(-Q^n)$ \u{is} a boundary of a connected $(n+1)$-dimensional space. (We consider only oriented manifolds; $-Q^n$ results from reversing orientation.) One says that $N^n$ and $Q^n$ are \u{cobordant} \cite{17}. In general, this will not solve our problem, because it will lead to the Witten-Yau paradox. However, if we choose $Q^n=N^n$, then we can represent $N^n$ as a connected infinity hypersurface by performing a topological identification. Unfortunately, this seems a somewhat arbitrary proceeding, because there are many other choices for $Q^n$ \- cobordant manifolds can be very different. We should ask whether the choice $Q^n=N^n$ can be motivated on \u{physical} grounds.

Before discussing this, let us consider the concept of \u{spin cobordism}. Let $\overline{M}^{n+1}$ be a compact manifold-with-boundary with interior $M^{n+1}$. If $M^{n+1}$ is a spin manifold, a given spin structure induces a spin structure on the boundary in a canonical way (\cite{12}, page 90). Now spin manifolds $N^n$ and $Q^n$ are \u{spin} cobordant if there exists a compact manifold-with-boundary $\overline{M}^{n+1}$ having $N^n+(-Q^n)$ as boundary, and having an interior with a spin structure that induces the given spin structures on $N^n$ and $-Q^n$. Clearly spin cobordism is the appropriate cobordism theory for physical applications. The spin cobordism equivalence classes in a given dimension form an abelian group, $\Omega^{spin}_n$. In low dimensions they are (\cite{12}, page 92):
\begin{align}
\Omega^{spin}_1&=\Z_2,&\Omega^{spin}_2&=\Z_2,\\
\Omega^{spin}_3&=0,&\Omega^{spin}_4&=\Z, \nonumber\\
\Omega^{spin}_5&=0,&\Omega^{spin}_6&=0,\nonumber\\
\Omega^{spin}_7&=0,&\Omega^{spin}_8&=\Z\oplus\Z.\nonumber
\end{align}

Let us apply these results. The two-dimensional case is instructive, and we begin with it. The group $\Omega^{spin}_2$ is generated by the torus $T^2$. This space has several spin structures, and one of them is not induced by any spin structure on the interior, $B^2\times S^1$; so $T^2$ with this spin structure is not a \u{spin} boundary (though of course it is a boundary as an \u{oriented} surface.) Now let $Q^2$ be spin cobordant to $T^2$, and set up a conformal field theory on $T^2+(-Q^2)$. As usual, we require stability, meaning that the scalar curvature \- the Gaussian curvature \- of $Q^2$ must be positive or zero. But if it is positive, the Gauss-Bonnet theorem identifies $Q^2$ as $S^2$, which is not spin cobordant to $T^2$ with this spin structure. Thus $Q^2$ is forced, by the stability condition, to be just another copy of $T^2$. We are thus invited to perform the appropriate identification and to study the CFT on $T^2$ by regarding it as an infinity hypersurface in a compact three-dimensional manifold.

This example gives us hope that the CFT stability condition might constrain the choice of $Q^n$ in higher dimensions, and in fact this is correct for the important case of $n=4$. Suppose that $N^4$ is a compact connected 4-manifold which admits a stable CFT but which is not a spin boundary. It is known that $\Omega^{spin}_4$ is generated by $K3$, so $N^4$ is spin cobordant to $m$ disjoint copies of $K3$, where $m$ could be negative but cannot, by hypothesis, be zero. Now the $\hat{A}$ genus is a spin corbordism invariant (\cite{12}. pages 92, 298) and since $\hat{A}(K3)=2$, we see that $\hat{A}(N^4)\not=0$. By the theorem of Lichnerowicz (\cite{12}, page 161) there is \u{no} metric of positive scalar curvature on $N^4$. Suppose that $Q^4$ is spin cobordant to $N^4$, so that $N^4+(-Q^4)$ is a spin boundary; then all of the above applies equally to $Q^4$.

The stability of the CFT on $N^4+(-Q^4)$ now demands that both $N^4$ and $Q^4$ admit metrics of zero scalar curvature. Now since $\hat{A}(N^4)$ and $\hat{A}(Q^4)$ are non-zero, both admit non-trivial harmonic spinors; but spinors on a compact spin manifold which are harmonic with respect to a scalar-flat metric must in fact be \u{parallel} (\cite{12}, page 161). The existence of non-trivial parallel spinors forces the spin holonomy group of a manifold to be \u{special}; it also forces the Ricci tensor to vanish \cite{18}. Now the spin holonomy groups of (not necessarily simply connected) compact Ricci-flat Riemannian spin manifolds have been classified \cite{19} and so we can conclude that the metrics on $N^4$ and $Q^4$ are either flat (which would contradict the non-vanishing of $\hat{A}(N^4)$ and $\hat{A}(Q^4)$) or of spin holonomy precisely $SU(2)$. (By ``precisely'', we mean that the full, global holonomy group (both linear and spin) is $SU(2)$, not just the identity component. There are four-manifolds with disconnected linear holonomy groups having $SU(2)$ as identity component, but these are \u{not} spin manifolds.) But every compact four-manifold of holonomy precisely $SU(2)$ is diffeomorphic to a finite number of copies of $K3$ (see \cite{8}, page 365). Since $\hat{A}(K3)=2$, we see that $m_1$ copies of $K3$ are spin cobordant to $M_2$ copies only if $m_1=m_2$, and so, since $N^4$ is connected, $Q^4$ must be diffeomorphic to it. Finally, the moduli space of Einstein metrics on $K3$ is connected (\cite{8}, page 366), and since the total scalar curvature is locally constant as a function on moduli space (\cite{8}, page 352), the $SU(2)$ metric on $Q^4$ can be deformed through scalar-flat metrics so that $Q^4$ is isometric to $N^4$. Regarding $N^4+(-Q^4)$ as the boundary of a compact manifold-with-boundary $\overline{M}^5$, we can search for an appropriate metric $g^M$ on $M^5$ such that $N^4+(-Q^4)$ is the conformal boundary (see below). Performing the identification as usual, we now have $N^4$ as an infinity hypersurface in a five-dimensional compact manifold, and we can begin to explore AdS/CFT for $N^4$, despite the fact that it is not a boundary.

In dimension 8, the situation is much more complex. The group $\Omega^{spin}_8=\Z\oplus \Z$ is generated by the quaternionic projective space $\h P^2$ and by any Joyce manifold $J_8$ of holonomy $Spin(7)$ (see \cite{20}). As $\h P^2$ is the symplectic homogeneous space $\dfrac{Sp(3)}{Sp(1)\times Sp(2)}$, it admits a metric of positive scalar curvature, and so, therefore, does any simply connected eight-manifold which is spin cobordant to it or to any finite number of copies of it (\cite{12}, page 299). On the other hand, there are scalar-flat metrics on the various candidates for $J_8$ and on many topologically distinct manifolds spin cobordant to $J_8$ (such as manifolds of linear holonomy $SU(4)\triangleleft\Z_2$) and to multiple copies of it (such as Calabi-Yau and hyperK\"{a}hler manifolds). Thus if $N^8$ is a non-boundary eight-manifold, there are many candidates for $Q^8$ and there is no good physical justification for selecting $Q^8=N^8$. Perhaps this indicates some kind of pathology afflicting AdS/CFT for conformal field theories on manifolds of dimension greater than seven.

\section{\large THE BRANE IN THE BULK}

Let $\hat{M}^5=S^1\times P^4$, where $P^4$ is a non-boundary four-manifold with a scalar-flat metric. We saw above that in fact $P^4$ is Ricci-flat, so if we use equation (4) to define a metric $g^M$ on $M^5$, then the Ricci tensor of $g^M$ satisfies, by equation (5) and (6),
\begin{align}
(R^M)^\theta_\theta&=-n,\\
(R^M)^i_j&=-n\delta^i_j+(n-1)\sin^2(\frac{\theta}{2})\delta^i_j.
\end{align}
Thus $M^5$ is not an Einstein manifold except ``near infinity''.

Now although we should be prepared ultimately to consider non-Einstein metrics on the bulk, we might prefer to begin with Einstein metrics and then to consider perturbations around them. One can of course do this for $S^4$ and for $S^1\times S^3$; but can one do it for infinities which are not boundaries? The following very remarkable theorem, which follows straightforwardly from results of Witten-Yau \cite{4} and Cai-Galloway \cite{5}, is relevant.
\begin{thm}[Witten-Yau-Cai-Galloway]
Let $\hat{M}^{n+1}$ be a compact $(n+1)$-dimensional manifold admitting an infinity hypersurface $N^n$ with respect to a Riemannian metric $g^M$ on the complement (which we assume to be orientable). Suppose that $N^n$ corresponds to a non-trivial element of the homology group $H_n(\hat{M}^{n+1},\Z)$ and that the induced conformal structure on $N^n$ is represented by a Yamabe metric of positive or zero scalar curvature. Then the equation $Ric(g^M)=-ng^M$ requires $g^M$ to be an \u{incomplete} metric.
\end{thm}

This result means that, in the case at hand, any attempt to force the bulk to be Einstein everywhere will merely cause the metric to develop some kind of pathology. [We require
all metrics to be differentiable, so incompleteness includes failures of differentiability.]
In physical terms, we can think of $Ric(g^M)=-ng^M$ as characterising the vacuum, and of the ``pathology'' as some kind of localised matter, \u{such as a brane}. The theorem then simply means that the presence of a non-boundary infinity hypersurface entails the existence of some such object in the bulk.

An example will be helpful. Let $P^4$ be a compact non-boundary four-manifold with a scalar-flat (hence, Ricci-flat) metric $g^P$. Let $\hat{M}^5=S^1\times P^4$, with $S^1$ now parametrised by $\theta$ in $(-\pi,\pi])$ (not 0 to $2\pi$ as before). Take $M^5$ to be $(S^1-\{0\})\times P^4$, let $\delta$ satisfy $0<\delta<\pi$, and define a metric $g^M_\delta$ by
\begin{align}
g^M_\delta&=\theta^{-2}(d\theta\otimes d\theta +g^P),&& \theta\in (0,\delta)\cup (-\delta,0), \\
&=h(\theta)(d\theta\otimes d\theta +g^P),&& \text{elsewhere.}
\end{align} 
Here $h(\theta)$ is a function which interpolates continuously and smoothly between the two ``$\theta^{-2}$ regions''. Clearly this manifold has a single, connected infinity hypersurface at $\theta=0$. Just as for the metric given by equation (4), however, we can make $\theta=0$ seem disconnected by changing coordinates :
\begin{equation*}
\theta=\begin{cases}\pi e^{-x},&x\geq0,0<\theta<\delta;\\
-\pi e^{x},& x<0,-\delta<\theta<0,\end{cases}
\end{equation*}
for the metric becomes $dx\otimes dx+\pi^{-2} e^{2\abs{x}}g^P$, with infinities apparently at $x=\pm \infty$. Notice the formal similarities to the Randall-Sundrum \cite{21} metric, which has $e^{-2\abs{x}}$ instead of $e^{2\abs{x}}$, and where $g^P$ would be flat, not just Ricci-flat. In fact, however, a simple calculation shows that the ``pseudo-Randall-Sundrum metric'' $dx\otimes dx + \pi^{-2}e^{2\abs{x}}g^P$ is an Einstein metric as long as $g^P$ is Ricci-flat \- it does \u{not} need $g^P$ to be flat (see \cite{8}, page 268). So we have 
\begin{equation}
Ric(g^M_\delta)=-4g^M_\delta, \qquad \theta\in (0,\delta)\cup(-\delta,0).
\end{equation}
That is, $g^M_\delta$ is precisely Einstein outside the immediate neighbourhood of $\theta=\pi$.

The Randall-Sundrum metric has a pathology at $x=0$ because of the absolute value function $\abs{x}$ in $e^{-2\abs{x}}$, due to the presence of a brane. Clearly the pseudo-Randall-Sundrum metric has the same property, and the WYCG theorem asserts that setting $\delta=0$ will cause $g^M_\delta$ to become incomplete at $\delta=\pi$. That is indeed the case, since the connection coefficient $\Gamma^\theta_{\theta\theta}$, for example, is discontinuous there. However, this shows that the pathologies required by the theorem can be rather mild \- one should not think of them as singularities, but rather as branes. If $\delta$ is not zero but extremely small, then equation (13) is satisfied everywhere in $M^5$ except in an extremely thin slice. Inside that slice, the metric is given by (12), and, turning the WYCG theorem around, we deduce that (13) is certainly \u{not} satisfied everywhere in the slice. Again, this is evidence that some kind of localised matter is present.

If we now relax the condition that $g^M$ be an Einstein metric, we can hope to learn
more about the nature of the matter whose existence is necessitated by this interpretation
of ``infinity''. Set 
\begin{equation}
Ric(g^M)=-ng^M+S(g^M),
\end{equation}
so that the tensor $S$ measures the failure of $g^M$ to be Einstein. In the Lorentzian case,
one could try to impose sign conditions on $S$ by means of the strong energy condition
\cite{10}, but that is not appropriate in the Euclidean regime. [The Euclidean counterpart
of a Lorentzian metric which satisfies the strong energy condition need not obey any sign
condition; consider for example the Euclidean version of a FRW dust metric, where the
Ricci curvature is unbounded both above and below.] Nevertheless, even in the Euclidean
case, $S$ does have non-negative eigenvalue functions \u{for certain kinds of matter}, 
such as scalar fields with positive potentials. Thus, the sign of $S$ can give general
information on the kind of matter which causes a given metric to be non-Einstein. The
following theorem is therefore relevant. [The proof is again a straightforward consequence 
of results of Cai and Galloway \cite{5}].
\begin{thm}[Cai-Galloway]
Let all conditions be as in the WYCG theorem, except that $Ric(g^M)=-ng^M$ is weakened to the condition that all of the eigenvalues of $S(g^M)$ decay no more slowly than inverse-quartically (see \cite{5} for details) towards infinity. Then \u{either} $g^M$ is incomplete \u{or} some eigenvalue function of $S(g^M)$ takes a value strictly less than $0$.
\end{thm}
[Notice that the metric (4) escapes the conclusions of this theorem because its Ricci
tensor does not tend to $-ng^M$ quickly enough.] Thus, for example, if the metric given by
(11) and (12) is forced to be complete by an appropriate choice of $h(\theta)$, then
$h(\theta)$ must be such that some eigenvalue of the Ricci tensor falls below $-4$, which
is the value outside the slice.

The above theorem leads us to ask: what kind of matter has an $S$-tensor with negative
eigenvalues? For example, can p-branes give rise to fields with such an $S$-tensor? The
answer is yes, as we now show. Consider the following action.
\begin{equation*}
\int [R+\lambda-\frac{1}{2}(\nabla\phi)^2-\frac{1}{2(p+2)!}e^{a\phi}F^2]dvol,
\end{equation*}
where $R$ is the scalar curvature, $\lambda$ is a constant to be chosen, $\phi$ is a scalar
field, $a$ is a constant, and $F$ is a $(p+2)$-form derived from a potential in the usual
way. Now p-brane solutions of the corresponding Euler-Lagrange equations are studied in
\cite{22}. A particularly interesting sub-class of \u{non-singular} solutions is obtained
when $\phi$ and $a$ can consistently be set equal to zero, so that only $F$ contributes to
$S$. This is possible [see \cite{22}, end of section 3.2] for
\begin{equation*}
(n,p) = (4,1), (5,1), (9,3),
\end{equation*}
that is, for strings in 5 and 6 dimensions, and for 3-branes in 10 dimensions. Selecting
$\lambda$ so that the coefficient of $g^M$ is $-n$ [as in (14)], we obtain in this case,
relative to a coordinate basis,
\begin{equation*}
S_{\mu\nu}=\frac{1}{2(p+1)!}[F_{\mu...}F_{\nu}^{...}-\frac{p+1}{(p+2)(n-1)}F^2g_{\mu\nu}],
\end{equation*}
[see \cite{22}, section 2.1], and so [recalling that the space is $(n+1)$-dimensional]
we have
\begin{equation*}
g^{\mu\nu}S_{\mu\nu}=\frac{(n-2p-3)F^2}{2(p+2)!(n-1)}.
\end{equation*}

Now for a 3-brane in 10 dimensions this vanishes, so $S$ is traceless. For a non-trivial
solution, this means that \u{some eigenvalue of S must indeed be negative somewhere}. For
a string in 5 dimensions, the trace is negative, while for a string in 6 dimensions it
vanishes; in both cases, $S$ is again forced to have eigenvalues which are negative somewhere.
For all of these solutions, $F$ decays towards infinity as [\cite{22},section 4] 
$1/r^{(n-p-1)}$, so $S$ decays at least as rapidly as $1/r^4$, as required by the above 
theorem. In short, the theorem is at least consistent with the hypothesis that some kind
of p-brane is present.

We can summarise as follows. The WYCG and Cai-Galloway theorems imply that the presence
of a topologically non-trivial infinity hypersurface imposes conditions on the geometry
of the bulk. We argue that those conditions suggest that the bulk has been contaminated
with ``branes or stringy impurities of some kind'', precisely as predicted in \cite{3}.

\section{\large CONCLUSION}

We have argued that the problem of formulating AdS/CFT for manifolds which are not boundaries suggests that we need an alternative framework, which does away with the need to deal with manifolds-with-boundaries. This framework, which uses compact manifolds with infinity hypersurfaces, may find uses even for cases where infinity \u{can} be represented as a boundary. For example, it could be interesting to investigate a CFT on the four-torus $T^4$ by thinking of it as a submanifold of $T^5$. Using the metric (11) above with $\delta=0$, we obtain an Einstein (in fact, a locally (Euclidean) AdS) space with a ``pseudo-Randall-Sundrum'' brane at the antipode to infinity.

In this work, we have followed the usual practice, relating the CFT to a gravitational theory in \u{one} more dimension. As string and $M$ theories are defined in 10 or 11 dimensions, however, this really just means that we are considering products, like $AdS_5\times S^5$, for the bulk. Presumably a generic bulk will not have this special structure, but a 10-dimensional manifold-with-boundary does not have a four-dimensional boundary. As a boundary cannot itself have a boundary, one cannot work stepwise down to four dimensions within the ``infinity as a boundary'' interpretation. By contrast, one can easily consider a four-dimensional submanifold in a generic compact 10-dimensional manifold endowed with a metric which puts that submanifold ``infinitely far'' from the points in the bulk. It would be interesting to develop AdS/CFT in this more general setting.

\end{document}